\documentclass[showpacs,pre,amssymb,amsmath,preprint]{revtex4}

\usepackage{psfrag}
\usepackage{amsmath}
\usepackage{amssymb}
\usepackage{textcomp}		% pour le micrometre de la table 4
\usepackage{array}		% pour la figure 10 !
\usepackage{multirow}		% pour le tableau 1
\usepackage{rotating}		% pour le texte vertical dans le tableau
\usepackage{floatflt}
\usepackage{textcomp}		% pour le micrometre de la table 4
\usepackage{color}

\numberwithin{equation}{section}

\newcommand{\be}{\begin{equation}}
\newcommand{\ee}{\end{equation}}
\newcommand{\bea}{\begin{eqnarray}}
\newcommand{\eea}{\end{eqnarray}}
\def \la{\label}

\def\({\left (}
\def\){\right )}
\def\]{\right]}
\def\[{\left[}
\def\<{\left <}
\def\>{\right>}

\newcommand{\bx}{\mathbf{x}}

\newcommand{\br}{\mathbf{r}}

\newcommand{\bk}{\mathbf{k}}

\newcommand{\cE}{\mathcal{E}}

\newcommand{\cS}{\mathcal{S}}

\newcommand{\cL}{\mathcal{L}}

\newcommand{\bcX}{\pmb{\mathcal{X}}}

\renewcommand{\d}{\mathrm{d}}

\newcommand{\Tr}{\mathrm{Tr}}
\newcommand{\e}{\mathrm{e}}

\newcommand{\sx}{\textsf{\textbf{x}}}
  
\begin{document}

  \title{Screened activity expansion for the pressure of a quantum multi-component plasma and 
	consistency with the local charge neutrality}

\author{A. Alastuey$^{1}$, V. Ballenegger$^2$ and D. Wendland$^{1,2}$ 
       \\[2mm]
       $^1${\small Laboratoire de Physique, ENS Lyon, UMR CNRS 5672}\\[-1mm]
        {\small 46 all\'ee d'Italie, 69364 Lyon Cedex 07, France}\\[-1mm]
              $^2${\small Institut UTINAM, Universit\'e de Bourgogne-Franche-Comt\'e, UMR CNRS 6213} \\[-1mm] 
							{\small 16, route de Gray, 25030 Besan\c{c}on Cedex, France}}

  \date{\today}
\begin{abstract}
We consider a quantum multi-component plasma made of point charged particles interacting \textit{via} the two-body 
Coulomb potential. Within the Feynman-Kac path integral representation of the system in terms of a classical gas of loops, 
we derive screened activity series for the pressure in the grand-canonical ensemble. The method is based on the Abe-Meeron summations which remove all 
long-range Coulomb divergences. Moreover, we show that the particle densities can be inferred from the diagrammatic series 
for the pressure, through partial differentiations with respect to suitable effective activities, consistently 
with the local charge neutrality. We briefly argue how these results can be used for including, in the equation of state at
moderately low densities, the contributions of recombined entities 
made with three or more particles.

\end{abstract}
\maketitle

%\tableofcontents

%===============================================

%===============================================
\section{Introduction}

Mayer diagrams have been introduced while ago in order to derive 
low-density expansions of equilibrium quantities for classical 
systems with short-range pair interactions. For charged fluids, every Mayer diagram 
diverges because of the long-range of Coulomb interactions. 
Abe and Meeron showed that such divergences can be removed \textit{via}
systematic summations of convolution chains built with the Coulomb interaction. The whole Mayer series is then 
exactly transformed into a series of so-called prototype graphs, with the same topological structure as the Mayer diagrams, but 
with effective bonds built with the familiar Debye potential in place of the bare Coulomb interaction. The contribution of each 
prototype graph is finite thanks to the screening  collective effects embedded in the Debye potential.

\bigskip

For a quantum Coulomb system, it turns out that Mayer-like diagrammatic series can be also introduced within its path integral representation 
in terms of an equivalent classical gas of loops~\cite{Ginibre1971,Cornu1996,Martin2003}.  Hence, the Abe-Meeron strategy can be extended 
to the gas of loops which interact via pairwise Coulomb-like interactions. However, the corresponding summations are carried out in the 
activity series which, contrarily to the case of classical point particles, cannot be transformed into 
particle-density series through the principle of topological reduction. The method was applied to the derivation of low-activity series for the 
one-body and two-body particle distribution functions in a first step, while term-by-term inversion of these series 
combined with thermodynamic identities provide particle-density expansions 
of the pressure in a second step. The corresponding equation of state has been calculated up to order $\rho^{5/2}$~\cite{AP1992}, and is 
fully consistent~\cite{ABE2015} with results obtained independently using different formalisms, like the 
pioneering work of Ebeling~\cite{Ebeling1967,KKER} based on the effective-potential method, charge-expansions within 
the standard many-body perturbation theory~\cite{DSKR1995} or field-theory calculations~\cite{BY2001}.
 
\bigskip

In this paper, we derive the screened activity-series for the pressure of a quantum multi-component plasma 
within the loop gas mapping. These series would provide a direct access to the equation of state, once the activities are eliminated 
in favor of the densities. However, such elimination requires to avoid the pitfalls related to the local charge neutrality.  
As proved by Lieb and Lebowitz~\cite{LiebLebowitz1972}, the particle densities calculated in the grand-canonical ensemble always 
satisfy the local charge neutrality, whatever the activities are. This remarkable property is exploited for reducing the number of 
prototype graphs which contribute to the pressure at a given order in the small activities. At the same time, the particle densities 
determined through suitable differentiations of this pressure, do satisfy the local charge neutrality. We briefly argue how the 
corresponding scheme would allow one to improve previous results for the equation of state, 
on the analytical and/or numerical side.  

\bigskip

The paper is organized as follows. In Section~\ref{sec:S2}, we recall the main results proved by Lieb and Lebowitz for a quantum 
Coulomb system made with $\cS$  components. As a consequence of the neutrality property in the grand-canonical ensemble, 
we infer that the pressure depends on $(\cS-1)$ independent combinations of the $\cS$ activities $\{ z_{\alpha} \}$. 
This leads to the introduction of a set of $(\cS-1)$ effective activities $\{ y_{i} \}$ defined by suitable transformations of 
the usual activities $\{ z_{\alpha} \}$. The pressure $P$ then only depends on the set $\{ y_{i} \}$ and on the temperature, while 
the particle densities obtained by suitable partial differentiations with respect to the $y_i$'s do satisfy local charge neutrality. 
The screened activity series for $P(T;\{ z_{\alpha} \})$ is derived in Section~\ref{sec:S3}. The Abe-Meeron summations are applied to the 
Mayer diagrams for the equivalent gas of classical loops. The pressure is then expressed as a series of prototype graphs where  
the usual points are replaced by loops, which are connected by screened bonds and carry dressed statistical weights. Within the 
degeneracy property exposed in Section~\ref{sec:S2}, we introduce a pseudo-neutrality prescription for the set $\{ z_{\alpha} \}$
which allows us to determine the small-$y$ expansion of $P(T;\{ y_i \})$ at a given order by keeping a finite number of 
prototype graphs. As illustrated in Section~\ref{sec:S4} for the simple cases $\cS=2$ and $\cS=3$, such expansion
of $P(T;\{ y_i \})$  gives access to the corresponding expansion of particle densities, which is consistent order by order with the 
charge neutrality. Concluding comments are given in Section~\ref{sec:S5}.

\section{Overall neutrality in the grand-canonical ensemble}\label{sec:S2}
\subsection{Quantum multi-component Coulomb system and the thermodynamic limit}

We consider a quantum multi-component plasma made of $\cS$ species of charged point particles enclosed 
in a box with volume $\Lambda$.   
The species index is denoted $\alpha$, $\{ \alpha=1,...,\cS \}$. Each particle of species $\alpha$ 
has a mass $m_\alpha$, while it carries a charge $e_\alpha$ and a spin $s_\alpha$. Each of them 
obeys to either Bose or Fermi statistics, according to the integer or half-integer value of $s_\alpha$ respectively.
In order to ensure thermodynamic stability, at least one species needs to be fermions~\cite{LiebLebowitz1972} and there are both positively and negatively
charged species. The species
$\alpha=1,...,\cS$ and the position $\bx$ of a given particle is denoted by the single notation 
$\sx=(\alpha,\bx)$. The total interaction potential $U(\sx_1,...,\sx_N)$ of $N$ particles 
is the sum of pairwise pure Coulomb interactions,
\be
\la{IX.QMG1}
U(\sx_1,...,\sx_N)= \sum_{i <j} V_{\rm C}(\sx_i,\sx_j) 
\ee
with 
\be
\la{IX.QMG2}
V_{\rm C}(\sx_i,\sx_j)=e_{\alpha_i} e_{\alpha_j} v_{\rm C}(|\bx_i - \bx_j|)
\ee
and $v_{\rm C}(r)=1/r$. The corresponding non-relativistic Coulomb Hamiltonian reads
\be
H_{N}=-\sum_{i=1}^{N}\frac{\hbar^2}{2m_{\alpha_i}} \Delta_i + U(\sx_1,...,\sx_N)
\la{IX.QMG3}
\ee
where $\Delta_i$ is the Laplacian with respect to position $\bx_i$. The nucleo-electronic plasma 
is an example of such multi-component system, where the negative point charge are electrons, while 
all positive point charges are nuclei.

\bigskip

As proved by Lieb and Lebowitz, the present quantum multi-component plasma has a well-behaved thermodynamic 
limit (TL), and all statistical ensembles become equivalent in this limit. In the grand-canonical 
ensemble the TL is defined by fixing the chemical potentials $\mu_\alpha$ of each species as well as the 
inverse temperature $\beta=1/(k_{\rm B}T)$, and letting $\Lambda \to \infty$. 
The grand-partition function $\Xi_{\Lambda}$ of the finite system reads
\be
\Xi_{\Lambda}= \Tr \exp[-\beta(H-\sum_{\alpha=1}^{\cS} \mu_{\alpha}N_{\alpha})] \; ,
\label{VI.48}
\ee
where the trace runs on all particle numbers, not only on neutral configurations. 
The grand canonical pressure
\begin{align}
P_{\Lambda}(T, \{ \mu_{\alpha} \})=\frac{k_{\rm B}T \ln \Xi_{\Lambda}}{|\Lambda|}
\label{VI.48bis}
\end{align}
has a  well-defined TL 
\begin{align}
P(T;\{ \mu_{\alpha} \})= k_{\rm B}T  \lim_{\rm TL}\frac{ \ln \Xi_{\Lambda}}{\Lambda} \; ,
\label{PressureTL}
\end{align}
which is a function of the chemical potentials and of the temperature.

\bigskip

It turns out that the non-neutral configurations
associated with $\sum_{\alpha=1}^{\cS} e_{\alpha}N_{\alpha} \neq 0$,
have such large positive Coulomb energies that they eventually do not contribute to the trace~(\ref{VI.48}) in the TL. 
As a consequence the system maintains charge neutrality in the bulk
\begin{equation}
\sum_{\alpha=1}^{\cS}e_{\alpha}\rho_{\alpha}=0
\label{VI.49}
\end{equation} 
where $\rho_{\alpha}$ is the particle density of species $\alpha$. Thanks to the thermodynamic identity
\be
\la{DensityChemicalPotential}
\rho_{\alpha}= \frac{\partial P}{\partial \mu_{\alpha}}(T;\{ \mu_{\gamma} \}) \; ,
\ee
charge neutrality~(\ref{VI.49}) can be recast as
\begin{equation}
\sum_{\alpha=1}^{\cS}e_{\alpha} \frac{\partial P}{\partial  \mu_{\alpha}} (T;\{ \mu_{\gamma} \})=0 \; .
\label{VI.50}
\end{equation}

\subsection{Consequence of the Lieb-Lebowitz theorem}

The identity~(\ref{VI.50}) is valid for any set $\{ \mu_{\gamma} \}$. This means that the pressure depends in fact on 
$(\cS-1)$ independent combinations of the chemical potentials, while there exist one combination which is irrelevant. This remarkable property 
can be easily understood by first considering the simplest case of a two-component system with $\cS=2$ and $e_1e_2 <0$. 
Let us express both $N_1$ and $N_2$ in terms 
of the total particle number $N=N_1+N_2$ and of the total charge $Q=e_1 N_1 + e_2 N_2$. A straightforward calculation provides
\begin{align}
\la{N1N2}
&N_1= -\frac{e_2}{(e_1-e_2)}N + \frac{Q}{(e_1-e_2)} \nonumber \\
&N_2= \frac{e_1}{(e_1-e_2)}N  - \frac{Q}{(e_1-e_2)} \; .
\end{align}
Accordingly, we obtain
\be
\la{LinearChemicalP}
\mu_1 N_1 + \mu_2 N_2= \[-\frac{e_2}{(e_1-e_2)}\mu_1 + \frac{e_1}{(e_1-e_2)} \mu_2 \] N + (\mu_1-\mu_2) \frac{Q}{(e_1-e_2)} \; .
\ee
After inserting this decomposition into the trace~(\ref{VI.48}), 
we see that the combination $\mu=(e_1 \mu_2 -e_2\mu_1)/(e_1-e_2)$ determines the total particle density 
$\rho = \lim_{TL}\langle N \rangle_{\rm GC}/\Lambda$ in the thermodynamic limit, while the combination
$\nu=\mu_1-\mu_2$ becomes irrelevant in the TL since the charge density $\lim_{TL} Q/\Lambda$ always vanish. 
Hence the pressure $P(T;\{ \mu_{\gamma} \})$ reduces to a function of only two variables 
instead of three, namely $P(T;\mu_1,\mu_2)=P(T;\mu)$. The particle densities are then expressed as 
\begin{align}
\la{DensityChemicalBis}
&\rho_1= \frac{\partial P}{\partial \mu}(T;\mu)\frac{\partial \mu}{\partial \mu_1}(\mu_1,\mu_2)
=-\frac{e_2}{(e_1-e_2)} \frac{\partial P}{\partial \mu}(T;\mu) \nonumber \\
&\rho_2= \frac{\partial P}{\partial \mu}(T;\mu)\frac{\partial \mu}{\partial \mu_2}(\mu_1,\mu_2)
=\frac{e_1}{(e_1-e_2)} \frac{\partial P}{\partial \mu}(T;\mu) \; .
\end{align}
Note that the expressions~(\ref{DensityChemicalBis}) do satisfy the neutrality property~(\ref{VI.49}), whatever the function 
$P(T;\mu)$ is. Moreover, the total particle density becomes
\be
\la{TotalParticleD}
\rho=\rho_1 + \rho_2= \frac{\partial P}{\partial \mu}(T;\mu) \; ,
\ee
in agreement with the above interpretation of $\mu$. 

\bigskip

For multi-component systems with three or more components, we can determine in a similar way the irrelevant combination of the 
chemical potentials by noting that 
\be
\la{LinearChemicalN}
\sum_{\alpha=1}^{\cS} \mu_{\alpha}N_{\alpha}= \pmb{\mu} \cdot \mathbf{N}
\ee
where $\pmb{\mu}$ and $\mathbf{N}$ are vectors in a $\cS$-dimensional space with respective components $\mu_\alpha$ and $N_\alpha$. 
Let $\mathbf{e}$ be the $\cS$-dimensional vector with components $e_\alpha$. If we decompose $\pmb{\mu}$ as 
\be
\la{DecompositionChemical}
\pmb{\mu}=\pmb{\mu}_\perp + \frac{\pmb{\mu} \cdot \mathbf{e}}{\mathbf{e}^2} \; \mathbf{e}
\ee
where $\pmb{\mu}$ is orthogonal to $\mathbf{e}$, we find that the scalar product~(\ref{LinearChemicalN}) 
reduces to 
\be
\la{LinearChemicalN}
\sum_{\alpha=1}^{\cS} \mu_{\alpha}N_{\alpha}= \pmb{\mu}_\perp \cdot \mathbf{N} + \frac{\pmb{\mu} \cdot \mathbf{e}}{|\mathbf{e}|^2} Q \; .
\ee
This implies that the linear combination $\sum_\alpha \e_{\alpha}\mu_{\alpha}/|\mathbf{e}|$ is not relevant. A relevant combination 
can be built by forming the scalar product $\mu \cdot \mathbf{e}_\perp$ where $\mathbf{e}_\perp$ is
orthogonal to $\mathbf{e}$. For $\cS=2$, there are two opposite vectors $\mathbf{e}_\perp$ with components $\pm (-e_2,e_1)$, so there is a single generic
relevant combination, namely $(-e_2 \mu_1 + e_1 \mu_2)$. Note that the multiplication of this generic combination by $1/(e_1-e_2)$ 
does provide the chemical potential $\mu=(-e_2 \mu_1 + e_1 \mu_2)/(e_1-e_2)$ which controls the total number of particles, as shown above.
For $\cS \geq 3$, there are $(\cS-1)$ independent vectors $\{ \mathbf{e}_\perp^{(1)},\mathbf{e}_\perp^{(2)},...,\mathbf{e}_\perp^{(\cS-1)} \}$
which form a basis of the $(\cS-1)$-dimensional plane orthogonal to  $\mathbf{e}$. They generate $(\cS-1)$ independent linear combinations of the 
chemical potentials which are relevant. Of course the choice of the basis $\{ \mathbf{e}_\perp^{(1)},\mathbf{e}_\perp^{(2)},...,\mathbf{e}_\perp^{(\cS-1)} \}$ 
remains arbitrary, as well as that of the $(\cS-1)$ relevant combinations of the $\mu_\alpha$'s. 

\bigskip

For further purposes, it is useful to translate the previous considerations in terms of the particle activities
\be
z_\alpha=(2s_\alpha+1)\frac{\e^{\beta\mu_{\alpha}}}{(2\pi \lambda_\alpha^{2})^{3/2}}\; ,
\la{ParticleActivity}
\ee
where $\lambda_\alpha=(\beta \hbar^2/m_\alpha)^{1/2}$ is the de Broglie thermal wavelength of the 
particles of species $\alpha$. The $(\cS-1)$ relevant linear combinations 
$\{ \mu \cdot \mathbf{e}_\perp^{(1)}, \mu \cdot \mathbf{e}_\perp^{(2)},...,\mu \cdot \mathbf{e}_\perp^{(\cS-1)} \}$ 
provide the corresponding $(\cS-1)$ independent combination of the particle activities, 
\be
\la{RelevantActivity} 
y_i = K_i(\{z_\alpha \}) \quad \text{with} \quad i=1,...,\cS-1 \; .
\ee
Each function $K_i(\{z_\alpha \})$ is a monomial in the $z_\alpha$'s, 
\be
\la{RelevantActivityPower} 
K_i(\{z_\alpha \})=\prod_{\alpha=1,...,\cS} z_\alpha^{\omega_\alpha^{(i)}} \; ,
\ee
where the respective dimensionless powers $\omega_\alpha^{(i)}$ are proportional to the components $e_{\perp,\alpha}^{(i)}$ of the
vector $\mathbf{e}_\perp^{(i)}$. The corresponding proportionality constant $C_i$, which is the same for 
all species $\alpha$ and a given $i$, can be always chosen such that 
\be
\la{RelevantActivityPowerBis} 
\sum_{\alpha=1,...,\cS} \omega_\alpha^{(i)} = 1 \; . 
\ee
This ensures that the new variables $y_i$ have all the dimension of an activity, \textit{i.e.} a density. Accordingly, the 
pressure is a function of the temperature and of the $(\cS-1)$ effective activities $y_i$, i.e. 
$P(T;\{ \mu_{\alpha} \})=P(T;\{ z_{\alpha} \})=P(T;\{ y_i \})$. The thermodynamical identity~(\ref{DensityChemicalPotential}) 
which provides the particle densities is then rewritten as 
\be
\la{DensityNewActivity}
\rho_{\alpha}= z_\alpha \sum_{i=1}^{\cS-1} \frac{\partial \beta P}{\partial y_i}(T;\{ y_j \}) 
\frac{\partial K_i}{\partial z_{\alpha}}(T;\{ z_{\gamma} \})\; .
\ee
The total local charge density then reads
\be
\la{ChargeDensityNewActivity}
\sum_\alpha e_\alpha \rho_{\alpha}=  \sum_{i=1}^{\cS-1} \frac{\partial \beta P}{\partial y_i}(T;\{ y_j \}) 
\sum_\alpha e_\alpha z_\alpha \frac{\partial K_i}{\partial z_{\alpha}}(T;\{ z_{\gamma} \})\; ,
\ee
and it indeed vanishes as it should since
\begin{multline}
\la{OrthogonalityProperty}
\sum_\alpha e_\alpha z_\alpha \frac{\partial K_i}{\partial z_{\alpha}}(T;\{ z_{\gamma} \})
=C_i K_i(T;\{ z_{\gamma} \}) \sum_\alpha e_\alpha e_{\perp,\alpha}^{(i)} \\
=C_i K_i(T;\{ z_{\gamma} \}) \mathbf{e} \cdot \mathbf{e}_\perp^{(i)}=0 \; .
\end{multline}

\section{Activity expansion of the pressure}	\label{sec:S3}

\subsection{The equivalent classical gas of loops}

The trace~(\ref{VI.48}) defining 
$\Xi_{\Lambda}$ can be expressed in the position and spin space, where the corresponding states have to be symetrized 
according to Bose or Fermi statistics. The corresponding sum involves both diagonal and off-diagonal 
matrix elements of $\exp(-\beta H_N)$ in position space. Diagonal matrix elements account for Maxwell-Boltzmann 
statistics, while off-diagonal matrix elements describe exchange contributions.
Within the Feynman-Kac representation, 
all the matrix elements of $\exp(-\beta H_N)$ in position space can be rewritten as 
functional integrals over paths followed by the particles. 
The off-diagonal matrix elements generate open paths. However all the open paths followed by 
the particles exchanged in a given cyclic permutation, can be collected into a close filamentous object, namely a loop $\cL$, also 
sometimes called polymer in the literature. 
Each contribution of a given spatial matrix element of $\exp(-\beta H_N)$ for a given set of particles, can be 
related to that of a classical Boltzmann factor for a given set of loops. The last non-trivial step consists in showing that the sum 
of all these contributions, namely $\Xi_{\Lambda}$, can be exactly recast as the grand-partition function of 
a classical gas of loops, namely~\cite{Ginibre1971,Cornu1996,Martin2003}
\be
\Xi_{\Lambda}=\Xi_{\Lambda,{\rm Loop}}=\sum_{k=0}^{\infty}\frac{1}{N\;!}\[ \prod_{i=1}^{N} \int_{\Lambda}D(\cL_{i})z(\cL_{i})\]
\e^{-\beta U(\cL_{1},\cL_{2},\ldots,\cL_{N})} \;,
\la{IX.QMG6}
\ee
where loop phase-space measure $D(\cL)$, loop fugacity $z(\cL)$, and total interaction potential 
$U(\cL_{1},\cL_{2},\ldots,\cL_{N})$ are defined as follows.

\bigskip
 
A loop $\cL$ located at $\bx$ containing $q$ particles of species $\alpha$, 
is a closed path $X(s)=\bx+\lambda_{\alpha}\bcX(s)$, parametrized by  an 
imaginary time $s$ running from $0$ to $q$ where $\bcX(s)$, the shape of the loop, is a Brownian bridge 
subjected to the constraints $\bcX(0)=\bcX(q)=\mathbf{0}$. 
The state of a loop, collectively noted $\cL=\{\bx, \chi\}$, is defined by its position $\bx$ together with an 
internal degree of freedom $\chi=\{\alpha, q,\bcX\}$, which includes its shape $\bcX$ as well as 
the number $q$ of exchanged particles of species $\alpha$. The loop phase-space measure $D(\cL)$ means summation over all these 
degrees of freedom,
\begin{equation}
\int_{\Lambda} D(\cL)\cdots=\sum_{\alpha=1}^{\cS}\sum_{q=1}^{\infty}\int_{\Lambda}\d \bx\int D_{q}(\bcX)\cdots \; .
\label{L.27}
\end{equation}
The functional integration over the loop shape $D_{q}(\bcX)$  is the normalized Gaussian measure for 
the Brownian bridge $\bcX(s)$ entirely defined by its covariance
\be
\int D_{q}(\bcX)\bcX_{\mu}(s_{1})\bcX_{\nu}(s_{2})=q\delta_{\mu\nu}\[\min\left(\frac{s_{1}}{q},\frac{s_{2}}{q}\right)-\frac{s_{1}}{q}\frac{s_{2}}{q}\] \; .
\la{IX.QMG7bis}
\ee

\bigskip

The loop activity reads
\be
z(\cL)=(2s_\alpha+1)\frac{\eta_\alpha^{q-1}}{q}\frac{\e^{\beta\mu_{\alpha}q}}{(2\pi q\lambda_\alpha^{2})^{3/2}}\e^{-\beta U_{\rm self}(\cL)} \; ,
\la{IX.QMG8}
\ee
where the factor $\eta_\alpha$ is related to the genuine symmetrization or 
anti-symmetrization of the position and spin states, namely $\eta_\alpha=1$ for bosons and $\eta_\alpha=-1$ for fermions.
Moreover, $U_{\rm self}(\cL)$ 
is the self-energy of the loop which is generated by the interactions between the exchanged particles,  
\begin{align}
U_{\rm self}(\cL)=\frac{e_\alpha^2}{2}\int_{0}^{q}\d s\int_{0}^{q}\d s'(1-\delta_{[s][s']})\tilde{\delta}(s-s')
v_{\rm C}(\lambda_\alpha \bcX(s)-\lambda_\alpha \bcX(s')) \;,
\label{IX.QMG9}
\end{align}
with the Dirac comb 
\begin{equation}
\tilde{\delta}(s-s')=\sum_{n=-\infty}^{\infty}\delta(s-s'-n)=\sum_{n=-\infty}^{{\infty}}\e^{2i\pi n(s-s')}=\delta(\tilde{s}-\tilde{s'}) \; .
\label{IX.QMG10}
\end{equation}
The Dirac comb ensures that particles only interact at equal times $s$ along their paths, as required by the Feynman-Kac formula, 
while the term $(1-\delta_{[s][s']})$ removes the contributions of self-interactions. 

\bigskip

Eventually, the total interaction potential $U(\cL_{1},\cL_{2},\ldots,\cL_{N})$ is a sum of pairwise interactions,
\be
\la{IX.QMG11}
U(\cL_{1},\cL_{2},\ldots,\cL_{N})=\frac{1}{2} \sum_{i \neq j} V(\cL_i,\cL_j)
\ee 
with 
\begin{align}
&V(\cL_i,\cL_j)=e_{\alpha_i} e_{\alpha_j} \int_{0}^{q_i}\d s_i\int_{0}^{q_j}\d s_j \tilde{\delta}(s_i-s_j)
v_{\rm C}(\bx_i+\lambda_{\alpha_i}\bcX_i(s_i)-\bx_j-\lambda_{\alpha_j}\bcX_j(s_j)) \; .
\label{IX.QMG12}
\end{align}
The loop-loop interaction $V(\cL_i,\cL_j)$ is generated by the interactions between any particle inside $\cL_i$ and any particle inside $\cL_j$. 
Like in formula~(\ref{IX.QMG9}), the Dirac comb~(\ref{IX.QMG10}) guarantees that interactions are taken at equal times along particle paths.

\bigskip

The introduction of the gas of loops is particularly useful at low densities, because the standard Mayer diagrammatic expansions, valid 
for classical systems with pairwise interactions, 
can be straightforwardly applied by merely replacing points by loops. However, as in the case of classical Coulomb systems, the Mayer-like 
diagrams for the loop gas are plagued with divergences arising from the large-distance behavior
\be
\la{IX.QMG28}
V(\cL_{i},\cL_{j}) \sim\frac{q_{i} e_{\alpha_{i}} q_{j} e_{\alpha_{j}}}{|\bx_{i}-\bx_{j}|} \;\quad \text{when} \; |\bx_{i}-\bx_{j}| \to \infty \; .
\ee
Note that such behavior is nothing but the Coulomb interaction between point charges, because the finite spatial extensions of both loops $\cL_i$ 
and $\cL_j$ can be neglected with respect to their relative large distance $|\bx_{i}-\bx_{j}|$. It has been shown that all these long-range 
divergences can be removed within a suitable extension of the Abe-Meeron summation process 
introduced long ago for classical Coulomb fluids. The method has been applied for both the one- and two-body distribution functions~\cite{Cornu1996,ABCM2003}.
 In the next Section, we derive the corresponding Abe-Meeron series for the pressure.

\subsection{Abe-Meeron like summations for the pressure}

Here, we will not detail the various counting processes which are identical to those arising in the classical case. Note that 
simplified presentations of the summation process are extensively described in Refs.~\cite{Cornu1996} and~\cite{ABCM2003}. 
The key staring point is the decomposition of the Mayer bond
\be
\la{MayerBond}
b_{\rm M}(\cL_k,\cL_l)= \exp(-\beta V(\cL_k,\cL_l)) -1 \; ,
\ee
into
\be
\la{DecompositionBond}
b_{\rm M}(\cL_k,\cL_l)=b_{\rm T}(\cL_k,\cL_l) + b_{\rm I}(\cL_k,\cL_l) \; ,
\ee
with the interaction bond
\be
\la{InteractionBond}
b_{\rm I}(\cL_k,\cL_l)=-\beta  V(\cL_k,\cL_l) \; 
\ee
and the truncated bond
\be
\la{TruncatedBond}
b_{\rm T}(\cL_k,\cL_l)= \exp(-\beta V(\cL_k,\cL_l)) -1 + \beta  V(\cL_k,\cL_l) \; .
\ee
A loop $\cL_l$ which is singly connected to an articulation loop $\cL_k$ is called and ending loop. 
Then, the corresponding Mayer bond $b_{\rm M}(\cL_k,\cL_l)$ is decomposed as 
\be
\la{DecompositionBondEnding}
b_{\rm M}(\cL_k,\cL_l)=b_{\rm TE}(\cL_k,\cL_l) + b_{\rm I}(\cL_k,\cL_l) + [b_{\rm I}(\cL_k,\cL_l)]^2/2\; ,
\ee
with the truncated ending bond
\be
\la{TruncatedEndingBond}
b_{\rm TE}(\cL_k,\cL_l)=\exp (-\beta V(\cL_k,\cL_l)) - 1 + \beta V(\cL_k,\cL_l)-(\beta V(\cL_k,\cL_l))^2/2 \; .
\ee
Then, after inserting the decompositions~(\ref{DecompositionBond}) and~(\ref{DecompositionBondEnding}) in every diagram of the Mayer series for the pressure $P$, 
we proceed to systematic summations of chain convolutions $b_{\rm I} \ast b_{\rm I} \ast...b_{\rm I} \ast b_{\rm I} $ made with arbitrary numbers $p$
of interaction bonds $b_{\rm I}$. The sum of $p=1,2,... \infty$ single convolution chains between two fixed loops $\cL_i$ and 
$\cL_j$ generate the quantum analogue $\phi(\cL_i,\cL_j)$ of the Debye potential, which reads~\cite{BMA2003} 
\be
\la{IX.QMG47}
\phi(\cL_{i},\cL_{j})= \int_{0}^{q_i}\d s_i\int_{0}^{q_j}\d s_j  
\; \psi_{\rm loop}(\bx_j+\lambda_{\alpha_{j}}\bcX_j(s_j)-\bx_i-\lambda_{\alpha_i}\bcX_i(s_i), s_i-s_j) \; ,
\ee
with
\be
\la{IX.QMG48}
\psi_{\rm loop}(\br,s)=  \sum_{n=-\infty}^\infty \exp(2 i \pi n s) \tilde{\psi}_{\rm loop}(\br,n) \; \; 
\ee
and
\be
\la{IX.QMG49}
\tilde{\psi}_{\rm loop}(\br,n)= \int \frac{ \d \bk}{(2\pi)^3} \exp(i \bk \cdot \br)   \frac{4 \pi }{k^2 + \kappa^2(k,n)} \; .
\ee
Note that $\tilde{\psi}_{\rm loop}(\br,n)$ has a structure analogous to the classical Debye form, 
except that an infinite number of frequency dependent screening factors $\kappa^{2}(k,n)$ occur, 
\be
\la{IX.QMG45a}
\kappa^2(k,n)=4\pi \beta \sum_\alpha \sum_{q=1}^\infty q e_{\alpha}^2 \int_0^q \d s \exp(2 i \pi n s) \int D_{q}(\bcX)
\exp(i \bk \cdot \lambda_{\alpha}\bcX(s)) z(\chi) \; .
\ee
The collective effects are embedded in these screening factors $\kappa^2(k,n)$, while the frequencies $2 \pi n$ are the analogues 
of the familiar Matsubara frequencies in the standard many-body perturbative series.

\bigskip

Similarly to the case of the Mayer diagrams for the one-body loop density, the summation of all convolution chains in the Mayer 
diagrams for the pressure can be expressed in terms of $\phi$, except in the single ring diagrams built with arbitrary numbers $p \geq 2$ of interaction 
bonds $b_{\rm I}$. In such diagrams made with $p$ black points (loops), the symmetry factor is $1/(2p)$, while in the single chain diagram 
connecting two different loops with $p$ intermediate black points and $(p+1)$ interaction bonds $b_{\rm I}$, the symmetry factor is $1$ for any $p$. 
After replacing each bare interaction $V$ by its decomposition over Matsubara frequencies, we find that the contribution to the pressure of a 
single ring made with $p$ black loops and $p$ bonds $b_{\rm I}=-\beta V$ reduces to
\be
\la{RingPp}
\frac{1}{2}\sum_{n=-\infty}^\infty \int \frac{\d \bk}{(2\pi)^3}\frac{1}{p} \left[\frac{-\kappa^2(k ,n)}{k^2}\right]^p \; .
\ee
The calculation is similar to that involved in the convolution chain and gives again raise to the screening factors $\kappa^{2}(k,n)$.
Now, the summation over $p$ of all ring contributions leads to a logarithmic function instead of the rational fraction $1/[k^2 + \kappa^{2}(k,n) ]$ for
the chain contributions, namely
\begin{equation}
\label{RingSum}
\beta P_{\rm R} = \frac{1}{2}\sum_{n=-\infty}^\infty  \int \frac{\d \bk}{(2\pi)^3} \left[
\frac{\kappa^2(k ,n)}{k^2} - \ln\left(1+ \frac{\kappa^2( k,n)}{k^2}\right) \right] \; .
\end{equation}

\bigskip

The summations for all the remaining diagrams are carried out as for the one-body density~(\ref{ABCM2003}). 
They generate the same screened bonds and the same dressed activities. The summations of single convolution chains between two 
fixed loops $\cL_i$ and $\cL_j$ made with arbitrary
numbers of intermediate loops and bonds $b_{\rm I}$ provides the Debye-like bond 
\be
\la{DebyeBond} 
b_{\rm D}(\cL_i,\cL_j)=-\beta e_{\alpha_i} e_{\alpha_j} \phi(\cL_i,\cL_j) \; .
\ee
The so-called Abe-Meeron bond,
\be
\la{AMBond} 
b_{\rm AM}(\cL_i,\cL_j) = \exp \( b_{\rm D}(\cL_i,\cL_j) \) - 1 - b_{\rm D}(\cL_i,\cL_j) \; .
\ee 
is obtained by summing more complex structures connecting 
the fixed pair $\cL_i$ and $\cL_j$ which may involve one bond $b_{\rm T}(\cL_i,\cL_j)$, one bond $b_{\rm I}(\cL_i,\cL_j)$ 
and one or more convolution chains made with arbitrary numbers $p \geq 2$
of interaction bonds $b_{\rm I}$. If $\cL_i$ is an ending loop, such summation provides the Abe-Meeron ending bond
\be
\la{AMEBond} 
b_{\rm AME}(\cL_i,\cL_j) = \exp \( b_{\rm D}(\cL_i,\cL_j) \) - 1 - b_{\rm D}(\cL_i,\cL_j) - [b_{\rm D}(\cL_i,\cL_j)]^2/2  \; .
\ee 

\bigskip
 
A first kind of dressed activity is obtained by summing all rings made with arbitrary numbers 
$p \geq 1$ of loops and $(p+1)$bonds 
$b_{\rm I}$ which can be attached to a  
given loop $\cL_i$. Inside each of these rings, the symmetry factor is now $1/2$ because of the particular role of the given attaching loop $\cL_i$. 
The summation over $p$ for single rings provides the ring sum
\begin{equation}
\la{XIIIRingSum}
I_{\rm R}(\cL_i) = \frac{1}{2}\left[ b_{\rm D}(\cL_i,\cL_i)-b_{\rm I}(\cL_i,\cL_i) \right] \;.
\end{equation}
Adding to the bare activity $z(\cL_i)$ the contributions of an arbitrary number of rings attached to $\cL_i$, we find
the dressed ring activity
\begin{equation}
\la{RingActivity}
z_{\rm R}(\cL_i)=z(\cL_i)\exp\left( I_{\rm R}(\cL_i) \right)\;.
\end{equation}  
Finally, the truncated dressed activity
\begin{equation}
\la{XIIIRingActivityWeak}
z_{\rm RT}(\cL_j)=z(\cL_i)\left( \exp\left( I_{\rm R}(\cL_j) \right)-1 \right)\;.
\end{equation}
also emerges for loops $\cL_i$ connected to the rest of the diagram by two bonds $b_{\rm I}$ and to which are attached one or more rings. 

\bigskip

The final screened Mayer series of the pressure reads
\begin{multline}
\la{ScreenedMayerP}
\beta P=  \beta P_{\rm R} + \int D(\cL) z(\cL) [e^{I_{\rm R}(\cL)}- I_{\rm R}(\cL)] \\
+ \sum_{{\cal P}} \frac{1}{S({\cal P})} 
\int \left[ \prod D(\cL) z^\ast(\cL) \right] 
\left[ \prod b^\ast \right]_{{\cal P}} \;.
\end{multline}
In this expression, the term next to the ring pressure $\beta P_{\rm R}$ is the contribution of the prototype graph made with a single black loop, 
and the sum in the second line is carried over all prototype graphs ${\cal P}$ made with $N \geq 2$ black points. 
Thanks to translation invariance, once the 
integration over $(N-1)$ black loops have been performed in ${\cal P}$, the result no longer depends on the position of the remaining black loop. The 
$1/\Lambda$ factor in the definition~(\ref{VI.48bis}) of the pressure of the finite system can then be absorbed in the TL by restricting the spatial 
integrations to the positions of $(N-1)$ loops in all graphs ${\cal P}$ and keeping one of them fixed. For instance, in the contribution of the 
simplest graph with one black loop $\cL$, it is understood that the integration $D(\cL)$ is carried out over all the internal degrees of 
freedom of loop $\cL$ except its position. The prototype diagrams ${\cal P}$ have the same topological structure 
as the genuine Mayer diagrams. They are simply 
connected and may contain articulation loops. Two loops can be connected by at most one bond $b^\ast$ and 
each loop carries a statistical weight $z^\ast$. There exist three possible bonds 
$b^\ast=b_{\rm D}, b_{\rm AM},b_{\rm AME}$ and three possible  
activities $z^\ast=z,z_{\rm R},z_{\rm RT}$. Their occurrence is determined by the specific 
rules listed below which avoid double counting. The sum in diagrammatic series~(\ref{ScreenedMayerP}) is carried out over all 
unlabeled prototype graphs with different topological structures: the contribution of a given ${\cal P}$ is 
calculated by labeling the $N$ field loops once for all. The symmetry factors $S({\cal P})$ are defined 
as usual, namely they are given by the number of permutations of those labeled field loops which leave the product of bonds 
and weights unchanged.

\bigskip

The activity $z^\ast(\cL_i)$ of a given field loop $\cL_i$ depends on the number of loops which are connected to it 
and on the corresponding bonds $b^\ast$, as follows :
\begin{itemize}
\item Ending field loop $\cL_i$ connected to a single loop $\cL_j$, which can be either a field or a root loop   
\begin{align}
\la{IX.QMG90}
&z^\ast(\cL_i)=z(\cL_i) \quad \text{for} \quad  b^\ast(\cL_i,\cL_j)=b_{\rm D}(\cL_i,\cL_j), b_{\rm AME}(\cL_i,\cL_j) \nonumber \\ 
&z^\ast(\cL_i)=z_{\rm RT}(\cL_i) \quad \text{for} \quad  b^\ast(\cL_i,\cL_j)=b_{\rm D}(\cL_i,\cL_j), b_{\rm AM}(\cL_i,\cL_j)  
\end{align}
\item Intermediate field loop $\cL_i$ in a convolution of two bonds $b^\ast(\cL_k,\cL_i)$ and $b^\ast(\cL_i,\cL_l)$ which connect $\cL_i$ 
to two loops $\cL_k$ and $\cL_l$, which can be either field or root loops  
\begin{align}
\la{IX.QMG91}
&z^\ast(\cL_i)=z_{\rm RT}(\cL_i) \quad \text{for} \quad  b^\ast(\cL_k,\cL_i)=b_{\rm D}(\cL_k,\cL_i) \; ,\; 
b^\ast(\cL_i,\cL_l)=b_{\rm D}(\cL_i,\cL_l) \nonumber \\ 
&z^\ast(\cL_i)=z_{\rm R}(\cL_i) \quad \text{for} \quad  (b^\ast(\cL_k,\cL_i),b^\ast(\cL_i,\cL_l)) \neq 
(b_{\rm D}(\cL_k,\cL_i),b_{\rm D}(\cL_i,\cL_l)) 
\end{align}
\item Field loop $\cL_i$ connected at least to three loops which can be either field or root loops  
\begin{equation}
\la{IX.QMG92}
z^\ast(\cL_i)=z_{\rm R}(\cL_i) \quad \text{for all} \quad  b^\ast
\end{equation}
\end{itemize}

\bigskip

The central quantity is the Debye bond 
$b_{\rm D}(\cL_i,\cL_j)=-\beta e_{\alpha_i} e_{\alpha_j} \phi(\cL_i,\cL_j)$. As shown in Ref.~\cite{BMA2002}, 
$\phi$ decays as $1/r^3$ at large distances $r$ 
between two loops. Thus bonds $b_{\rm AM}$ and $b_{\rm AME}$ decay respectively as $1/r^6$ and $1/r^9$, and they are integrable. 
The bond $b_{\rm D}$ decays as $\phi$ itself, \textit{i.e.} as $1/r^3$, which is at the border line for integrability. 
Accordingly, the graphs with ending loops connected to the rest of the diagram by bonds $b_{\rm D}$ have to be dealt with some care. 
In fact, since the corresponding weights of the ending loops, $z$ or $z_{\rm RT}$ are even functions of the loop shapes $\bcX(s)$, if we proceed first to
functional integrations over such shapes, then the $1/r^3$-algebraic tails vanish since their amplitudes are odd functions of $\bcX(s)$~\cite{BMA2002}.
Within this recipe, every prototype graph provides a finite contribution, as expected.

\subsection{Relation to the activity expansion of the species densities}

The screened activity expansion of the loop density can be readily inferred from 
the corresponding expansion~(\ref{ScreenedMayerP}) of the pressure by using 
\be
\la{FunctionalDerivativeP}
\rho(\cL_{\rm a})=z(\cL_{\rm a}) \frac{ \delta \beta P}{\delta z(\cL_{\rm a})} \; .
\ee
The functional derivative of each prototype diagram ${\cal P}$ is calculated by either whitening a black loop $\cL$
with weight $z(\cL)$ into the root loop $\cL_{\rm a}$ with weight $z(\cL_{\rm a})$ or by taking the functional derivative with 
respect to $z(\cL_{\rm a})$ of $\beta P_{\rm R}$ and $b_{\rm D}(\cL_i,\cL_j)$, namely
\be
\la{FDRingP}
z(\cL_{\rm a}) \frac{ \delta \beta P_{\rm R}}{\delta z(\cL_{\rm a})}= z(\cL_{\rm a}) I_{\rm R}(\cL_{\rm a})
\ee 
and 
\be
\la{FDphi}
z(\cL_{\rm a}) \frac{ \delta b_{\rm D}(\cL_i,\cL_j)}{\delta z(\cL_{\rm a})} = z(\cL_{\rm a})b_{\rm D}(\cL_i,\cL_{\rm a})b_{\rm D}(\cL_{\rm a},\cL_j) \; .
\ee
Note that the functional derivatives of the dressed activities and of the other bonds, which can be all expressed in terms of 
$b_{\rm D}$, are then obtained by using Eqn.~(\ref{FDphi}). This calculation provides 
\be
\la{IX.QMG88}
\rho(\cL_{\rm a}) = z_{\rm R}(\cL_{\rm a})  \sum_{{\cal P}_{\rm a}} \frac{1}{S({\cal P}_{\rm a})} 
\int \left[ \prod D(\cL) z^\ast(\cL) \right] 
\left[ \prod b^\ast \right]_{{\cal P}_{\rm a}} \; . 
\end{equation}
which can be also obtained by a direct Abe-Meeron summation of the Mayer diagrammatic series for the loop density~\cite{ABCM2003}.
The insertion of these series into the identity
\be
\la{ParticleDensityLoop}
\rho_{\alpha_{\rm a}} = \sum_{q_{\rm a}=1}^{\infty}\int D_{q}(\bcX_{\rm a}) \; \rho(\cL_{\rm a}) \; .
\ee
leads to required activity expansion of the species density.

\bigskip

The  prototype diagrams ${\cal P}_{\rm a}$  have the same topological structure 
as the genuine Mayer diagrams for the loop density. They are built with the root (white) loop $\cL_{\rm a}$ 
and $N$ black loops $\cL$ (N=0,1,...), simply 
connected and may contain articulation loops. Two loops can be connected by at most one bond $b^\ast$ and 
each black loop carries a statistical weight $z^\ast$. As for the graphs ${\cal P}$ in the series~(\ref{ScreenedMayerP}), 
there are again three possible bonds 
$b^\ast=b_{\rm D}, b_{\rm AM},b_{\rm AME}$ and three possible  
activities $z^\ast=z,z_{\rm R},z_{\rm RT}$, with the same occurrence 
rules. The sum $\sum_{{\cal P}_{\rm a}}...$ is carried out over all 
unlabeled prototype graphs with different topological structures. The symmetry factors $S({\cal P}_{\rm a})$ are defined 
as usual, like $S({\cal P})$.

\subsection{The pseudo-neutrality condition}

Let us consider a given graph ${\cal P}$ in the series~(\ref{ScreenedMayerP}), and the Debye dressed graph ${\cal P}_{\rm DD}$ obtained by 
adding a black loop $\cL$ with weight $z(\cL)$ connected to ${\cal P}$ \textsl{via} a single bond $b_{\rm D}(\cL,\cL')$ 
where $\cL'$ is a black loop inside ${\cal P}$. In the low activity limit, the 
potential $\phi$ reduces to its classical Debye counterpart~\cite{BMA2002}, so 
\be
\la{DebyeBondLowA}
b_{\rm D}(\cL,\cL') \sim -\beta q_\alpha e_\alpha q_{\alpha'}e_{\alpha'} \frac{\exp(-\kappa_z |\bx-\bx'|  )}{|\bx-\bx'|}
\ee
with 
\be
\la{KappaLowA}
\kappa_z^2=\kappa^2(0,0) \sim 4 \pi \beta \sum_\gamma  e_\gamma^2 z_\gamma \; .
\ee
At leading order in the small activities, the contribution of the graph ${\cal P}_{\rm DD}$ is obtained by keeping only the loop 
$\cL$ made with a single particle, i.e. $q_\alpha=1$, while the bond $b_{\rm D}(\cL,\cL')$ is replaced by its classical Debye 
expression~(\ref{DebyeBondLowA}). The corresponding leading contribution of ${\cal P}_{\rm DD}$ 
then reduces to that of graph ${\cal P}$ multiplied 
by 
\begin{multline}
\la{DebyeBondLowAbis}
\int D(\cL) z(\cL) \; b_{\rm D}(\cL,\cL') \sim -\beta  q_{\alpha'}e_{\alpha'} \sum_\alpha e_\alpha z_\alpha 
\int \d \bx \frac{\exp(-\kappa_z |\bx-\bx'|  )}{|\bx-\bx'|} \\
=-\frac{4 \pi \beta  q_{\alpha'}e_{\alpha'}}{\kappa_z^2} \sum_\alpha e_\alpha z_\alpha \; .
\end{multline}

\bigskip

According to the low-activity estimation~(\ref{DebyeBondLowAbis}), the contribution of ${\cal P}_{\rm DD}$ has the same order as 
that of ${\cal P}$ for arbitrary sets  $\{ z_\alpha\}$ of particle activities. In other words, in order to compute the pressure at 
a given order for such sets, one would have to keep an infinite number of graphs in the series~(\ref{ScreenedMayerP}). However, it turns out 
that the remarkable property of the function $P(T;\{ z_{\alpha} \})$ exposed in Section~\ref{sec:S2} allows us to circumvent this drawback. 
Indeed, since $P(T;\{ z_{\alpha} \})$ only depends on $(\cS -1)$ independent combinations of the activities
$y_1= K_1(\{ z_{\alpha} \}), y_2= K_2(\{ z_{\alpha} \}),..., y_{\cS -1}=K_{\cS -1}(\{ z_{\alpha} \}) $, the function 
$P(T;\{ y_{i} \})$  can be determined, without 
any loss of generality, by fixing an arbitrary relation between the $z_\alpha$'s. Hence, it is particularly useful to fix once for all 
the so-called pseudo-neutrality condition
\be
\la{PseudoNeutrality}
\sum_\alpha e_\alpha z_\alpha=0 \; .
\ee
Then, the leading contribution of ${\cal P}_{\rm DD}$ has an higher order than  
that of ${\cal P}$. At a given order in small activities, only a finite number of graphs ${\cal P}$ 
now have to be kept in the series~(\ref{ScreenedMayerP}). It then remains to invert the $\cS$ independent relations
\begin{align}
\la{ActivityInversion}
& y_1= K_1(\{ z_{\alpha} \}) \; ; \; y_2= K_2(\{ z_{\alpha} \}) \; ; \;...\; ; \; y_{\cS -1}=K_{\cS -1}(\{ z_{\alpha} \}) \nonumber \\
& \sum_\alpha e_\alpha z_\alpha=0 \; ,
\end{align}
in order to obtain the corresponding small-$y$ expansion of the pressure $P(T;\{ y_{i} \})$. 
Such simplification is crucial for further calculations of particle densities 
which do satisfy overall neutrality as exposed in the next Section~\ref{sec:S4}.

\section{Derivation of the particle densities consistent with the local charge neutrality}
\label{sec:S4}

Once the small-$y$ expansion of $P(T;\{ y_{i} \})$ has been obtained, the corresponding expansion 
of the particle densities is readily calculated via the identity~(\ref{DensityNewActivity}). The 
corresponding set $\{ \rho_\alpha \}$ of particle densities do automatically satisfy the local 
charge neutrality as shown in Section~\ref{sec:S2}. This is illustrated here for the two-component plasma by considering the 
first two terms of activity or density expansions. We also briefly introduce the case of a three component plasma.

\subsection{Two-component plasma}

There is a single vector $\mathbf{e}_\perp^{(1)}=\mathbf{e}_\perp$ with components $(-e_2,e_1)$. The powers $(\omega_1,\omega_2)$ 
in the function~(\ref{RelevantActivityPower}) which defines the effective activity 
\be
\la{NewActivityTCP}
y=z_1^{\omega_1} z_2^{\omega_2}
\ee
are 
\be
\la{PowersNewActivityTCP}
\omega_1= -\frac{e_2}{e_1-e_2} \quad \text{and} \quad \omega_2= \frac{e_1}{e_1-e_2} \; .
\ee
The expression~(\ref{DensityNewActivity}) for each species density becomes
\be
\la{ParticleDNA}
\rho_1= \omega_1 \; y \frac{\partial \beta P}{\partial y}(T; y)  \quad \text{and} \quad \rho_2= \omega_2 \;  y \frac{\partial \beta P}{\partial y}(T; y) \; .
\ee
Note that the charge density $(e_1 \rho_1 + e_2 \rho_2 )$ indeed vanishes, while the total particle density is 
\be
\la{TotalDNA}
\rho= \rho_1 + \rho_2=  y \frac{\partial \beta P}{\partial y}(T; y)  \; .
\ee

\bigskip

Within the pseudo-neutrality condition~(\ref{PseudoNeutrality}), the expression of both $z_1$ and $z_2$ in terms of $y$ is readily 
calculated as 
\be
\la{InversionZY}
z_1= (\omega_1/\omega_2)^{\omega_2} y \quad \text{and} \quad z_2= (\omega_2/\omega_1)^{\omega_1} y
\ee
The small-activity expansion of the pressure 
inferred from the screened Mayer series~(\ref{ScreenedMayerP}) then reads
\be
\la{LAExpansionP}
\beta P = z_1 + z_2 + \frac{\kappa_z^{3}}{12 \pi} +...
\ee
with $\kappa_z=[4\pi \beta (e_1^2z_1 + e_2^2z_2 )]^{1/2}$, while the terms left over are $o(z^{3/2})$. Accordingly the small-$y$ expansion of 
the function $\beta P(T,y)$ is
\be
\la{LNAExpansionP}
\beta P = \omega_1^{-\omega_1}\omega_2^{-\omega_2} \;  y + \frac{\kappa_y^{3}}{12 \pi} + o(y^{3/2})
\ee
with $\kappa_y=[4\pi \beta (e_1^2 (\omega_1/\omega_2)^{\omega_2}+ e_2^2 (\omega_2/\omega_1)^{\omega_1} ) y]^{1/2}$. 
Inserting the expansion~(\ref{LNAExpansionP}) into the identities~(\ref{ParticleDNA}), we find
\be
\la{ParticleDNABis}
\rho_1= \frac{\omega_1^{\omega_2}}{\omega_2^{\omega_2}} \; y + \frac{\omega_1}{8 \pi} \kappa_y^{3} + o(y^{3/2}) \quad \text{and} \quad
\rho_2= \frac{\omega_2^{\omega_1}}{\omega_1^{\omega_1}} \; y + \frac{\omega_2}{8 \pi} \kappa_y^{3} + o(y^{3/2})  \; ,
\ee
which can be recast in terms of the activities $z_1$ and $z_2$ as 
\be
\la{ParticleDA}
\rho_1= z_1 + \frac{\omega_1}{8 \pi} \kappa_z^{3}  + o(z^{3/2}) \quad \text{and} \quad
\rho_2= z_2 + \frac{\omega_2}{8 \pi} \kappa_z^{3} + o(z^{3/2})  \; .
\ee
On another hand, the particle densities can be also calculated by inserting the screened activity expansion~(\ref{IX.QMG88}) of the loop
density into the identity~(\ref{ParticleDensityLoop}). Within the pseudo-neutrality condition~(\ref{PseudoNeutrality}), the first two terms of order 
$O(z)$ and $O(z^{3/2})$ are obtained by keeping two diagrams: (i) the diagram made with a single root (white) loop $\cL_{\rm a}$ (ii) the diagram 
made with loop $\cL_{\rm a}$ connected to a single black loop by a Debye bond $b_{\rm D}$.  
This allows us to exactly recover the expressions~(\ref{ParticleDA}), as it should.

\bigskip

We stress that a direct calculation of $\rho_1$ and $\rho_2$ by taking the partial derivatives with respect to $z_1$ 
and $z_2$ respectively of the expression~(\ref{LAExpansionP}), as if $z_1$ and $z_2$ were independent variables, would lead 
to wrong expressions, which in particular do not satisfy the local charge neutrality. In particular, if the leading terms of 
order $O(z)$ would coincide with the ideal contributions in the correct expansions~(\ref{ParticleDA}), the terms of order 
$o(z^{3/2})$ would differ of their exact counterparts for a charge-asymmetric system with $\omega_1 \neq \omega_2$. 
If we pursue the expansions beyond the terms $O(z^{3/2})$, the direct calculation of the densities would be wrong at sufficiently high
orders, even for charge-symmetric systems. This sheds 
light on the importance of expressing the pressure in terms of the new activity $y$, in order to derive expansions consistent 
with the local charge neutrality.  

\bigskip

The hydrogen plasma is an example of charge-symmetric two-component plasma, made with protons ($e_1=e$) and electrons $e_2=-e$. Then, the powers $\omega_1$
and $\omega_2$ are identical, $\omega_1=\omega_2=1/2$, so we find $y=(z_1z_2)^{1/2}$. The helium plasma 
is an example of charge-asymmetric two-component plasma, made with $\alpha$ nuclei ($e_1=2e$) and electrons $e_2=-e$. The corresponding powers are 
$\omega_1=1/3$ and $\omega_2=2/3$, while the effective activity reduces to $y=z_1^{1/3} z_2^{2/3}$.

\subsection{Three component plasmas}

For three component plasmas, the two effective activities $y_1$ and $y_2$ are defined by choosing two independent vectors
$\{ \mathbf{e}_\perp^{(1)},\mathbf{e}_\perp^{(2)}\}$ orthogonal to $\mathbf{e}$. For fixing ideas, let us consider the case of
the hydrogen-helium mixture made with protons ($e_1=e$), $\alpha$-nuclei ($e_2=2e$) and electrons ($e_3=-e$). Since the components of 
$\mathbf{e}$ are $e(1,2,-1)$, we can choose $\mathbf{e}_\perp^{(1)}=e(1,0,1)$ and $\mathbf{e}_\perp^{(2)}=e(0,1,2)$. 
The corresponding effective activities are
\be
\la{NewActivity3CP}
y_1=z_1^{1/2} z_3^{1/2} \quad \text{and} \quad y_2=z_2^{1/3} z_3^{2/3} \; .
\ee
The identity~(\ref{DensityNewActivity}) provides the species densities 
\begin{align}
\la{ParticleDNA}
& \rho_1= \frac{y_1}{2} \frac{\partial \beta P}{\partial y_1}(T; y_1,y_2) \nonumber \\
&\rho_2= \frac{y_2}{3} \frac{\partial \beta P}{\partial y_2}(T; y_1,y_2) \nonumber \\
&\rho_3= \frac{y_1}{2} \frac{\partial \beta P}{\partial y_1}(T; y_1,y_2) 
+ \frac{2y_2}{3} \frac{\partial \beta P}{\partial y_2}(T; y_1,y_2) \; .
\end{align}
The local charge neutrality, $e \rho_1 + 2 e\rho_2 -e \rho_3=0$, is satisfied, while the total particle density 
reads 
\be
\la{TotalD3CP}
\rho = \rho_1+ \rho_2 +\rho_3= y_1 \frac{\partial \beta P}{\partial y_1}(T; y_1,y_2) 
+y_2 \frac{\partial \beta P}{\partial y_2}(T; y_1,y_2) \; .
\ee

\bigskip

Like in the two-component case, we can compute the small-$z$ expansion of the particle densities by 
using in the identities~(\ref{ParticleDNA}) the small-$y$ expansion of the pressure inferred from the screened 
activity series~(\ref{ScreenedMayerP}). Again, and unsurprisingly, we check explicitly 
that the lowest-order terms in  this small-$z$ expansion 
are exactly retrieved by using the screened activity series~(\ref{IX.QMG88}) for the loop density combined with the pseudo-neutrality 
condition~(\ref{PseudoNeutrality}).

\section{Concluding comments}
\label{sec:S5}

The various tools presented here should be quite useful for improving the equation of state of various quantum plasma at moderately 
low densities $\rho$. If the well-known virial expansions up to order $\rho^{5/2}$ provide an accurate description of the almost 
fully ionized regime observed at sufficiently low densities, recombination processes into entities made with three or more particles may
become important when the density increases. In such situations, where the density still remains relatively small, the 
corresponding contributions can be easily identified in 
the screened activity series for the pressure in terms of a few simple prototype graphs made with three or more loops. This 
defines a finite cluster function $Z_{\cE}$ for any entity (chemical species) $\cE$. 
The function $Z_{\cE}$ has to be expressed in terms of the effective activities
$\{ y_{i} \}$, in order to determine, consistently with charge neutrality, the corresponding contributions to the particle densities. 
Once all the relevant physical contributions have been calculated along similar lines, the expressions of the particle densities 
in terms of the effective activities are inverted. This provides the required equation of state in terms of particle densities, which 
correctly accounts, in a non-perturbative way, for the emergence of recombined entities. 

\bigskip

The physical content of the above scheme 
is very close to that of the celebrated ACTEX expansion introduced by Rogers~\cite{Rogers1974}. However, we stress that the cluster functions 
$Z_{\cE}$ are defined within systematic prescriptions which avoid double counting problems. Moreover they properly account for the collective 
screening effects which ensure their finiteness, without introducing ad-hoc regularisations as in the phenomenological 
Planck-Larkin partition functions. Note that the the cluster functions
$Z_{\cE}$ were first introduced within the screened activity series for the particle densities~\cite{ABCM2003}. It was shown that their 
zero-density limit leads to a natural definition of the bare partition function of chemical species in the vacuum.

\bigskip

Eventually, let us mention two applications of the previous general scheme. First, for 
the hydrogen plasma, atoms ${\rm H}$ and molecules ${\rm H}_{2}$ successively emerge when the density increases 
starting from the very dilute regime where all protons and electrons are ionized. In order to describe the 
cross-over regimes between the ionized, atomic and molecular phases, one has to first determine properly the various 
cluster functions describing atoms ${\rm H}$, ions  ${\rm H}_2^+$,  ${\rm H}^-$ and molecules ${\rm H}_{2}$ as
functions of the effective activity $y=(z_{\rm p}z_{\rm e})^{1/2}$. The accurate computation of the cluster functions involving 
more than two particles, \textit{i.e.} $Z_{{\rm H}_2^+}$, $Z_{{\rm H}^-}$ and $Z_{{\rm H}_{2}}$ require the introduction 
of suitable quantum Monte Carlo techniques~\cite{WBA2014}. Within a suitable double zero-density and zero-temperature
limit which defines a partially ionized atomic phase, exact asymptotic expansion beyond the familiar Saha theory~\cite{Saha1920}
have been already derived~\cite{ABCM2008}. 

\bigskip

For the hydrogen-helium mixture at low densities, the presence of helium atoms ${\rm He}$, as well as ions like ${\rm H}^-$ or ${\rm H}_2^+$, 
requires to consider prototype graphs made three loops. Once the corresponding cluster functions
$Z_{\rm He}$, $Z_{{\rm H}^-}$ and $Z_{{\rm H}_2^+}$ have been computed within the same 
quantum Monte Carlo techniques~\cite{WBA2014} as above, it remains to express them
in terms of the effective activities $(y_1,y_2)$ 
~(\ref{NewActivity3CP}) with $z_1=z_{\rm p}$, $z_2=z_{\alpha}$ and $z_3=z_{\rm e}$. A partial account of the corresponding calculations 
along the Sun adiabat is given in Refs.~\cite{WendlandThesis,BAW2018}.

\section*{Acknowledgements}

Financial support from the CNRS (contract 081912) and from the Conseil r\'egional de Franche-Comt\'e (contract 362887) are gratefully acknowledged. The calculations were run on computers from the Institute UTINAM of the University de Franche-Comt\'e, supported by the R\'egion Franche-Comt\'e and the Institut des Sciences de l'Univers (INSU).

\end{document}